\documentclass[prl,twocolumn, superscriptaddress]{revtex4-2}
\usepackage{xparse}
\usepackage{braket}
\usepackage{mathtools}
\usepackage{appendix}
\usepackage{graphicx}
\usepackage{dcolumn}
\usepackage{bm}
\usepackage{xcolor}
\usepackage[normalem]{ulem}
\usepackage{natbib}
\usepackage{array}
\usepackage{makecell}
\usepackage{multirow}
\begin{document}
\title{Natural Quantization of Neural Networks}
\author{Richard Barney}
\author{Djamil Lakhdar-Hamina}
\author{Victor Galitski}
\affiliation{Joint Quantum Institute, Department of Physics, University of Maryland, College Park, Maryland, USA}
\begin{abstract}
We propose a natural quantization of a standard neural network, where the neurons correspond to qubits and the activation functions are implemented via quantum gates and measurements. The simplest quantized neural network corresponds to applying single-qubit rotations, with the rotation angles being dependent on the weights and measurement outcomes of the previous layer. This realization has the advantage of being smoothly tunable from the purely classical limit with no quantum uncertainty (thereby reproducing the classical neural network exactly) to a quantum case, where superpositions introduce an intrinsic uncertainty in the network. We benchmark this architecture on a subset of the standard MNIST dataset and find a regime of ``quantum advantage,'' where the validation error rate in the quantum realization is smaller than that in the classical model. We also consider another approach where quantumness is introduced via weak measurements of ancilla qubits entangled with the neuron qubits. This quantum neural network also allows for smooth tuning of the degree of quantumness by controlling an entanglement angle, $g$, with $g=\frac\pi 2$ replicating the classical regime. We find that validation error is also minimized within the quantum regime in this approach. We also observe a quantum transition, with sharp loss of the quantum network's ability to learn at a critical point $g_c$. The proposed quantum neural networks are readily realizable in present-day quantum computers on commercial datasets. 
\end{abstract}

\maketitle

Recent advances in the fields of quantum computing and classical machine learning have fueled the growth of a new area of study examining the intersection of these two approaches to computation---quantum machine learning. Classical artificial neural networks (NNs) have demonstrated remarkable success across a wide range of tasks once thought to require human intelligence, such as complex classification tasks~\cite{Krizhevsky2012,Li2014}, natural language processing~\cite{Hirschberg2015,Otter2021,Lauriola2022}, protein structure prediction~\cite{Senior2020}, and materials data science~\cite{Choudhary2022,Butler2018}. 
At the same time, steady progress has been made in quantum computation~\cite{Gill2022}, which aims to take advantage of the fundamentally quantum aspects of nature, such as superposition of states, interference, and entanglement, using them as resources in computational tasks~\cite{Shor1994,Grover1996}. 

Naturally, there is interest in combining these two domains. Artificial neural network models that utilize quantum resources are called quantum neural networks (QNNs), and they have received significant attention in recent years~\cite{Jaderberg2022,Cerezo2022,Schuld2014,Abbas2021,Beer2020,Jeswal2019,Kwak2021}. Several formulations of QNNs have been proposed, such as parameterized quantum circuits~\cite{Benedetti2019,Cerezo2021,Stoudenmire2016,Mitarai2018}, quantum Hopfield networks~\cite{Rebentrost2018,Rotondo2018,Xie2024}, quantum Boltzmann machines~\cite{Amin2018,Zoufal2021,Bangar2025}, quantum perceptrons~\cite{Lewenstein1994,Kapoor2016,daSilva2016,Torrontegui2019}, and quantum convolutional neural networks~\cite{Cong2019,Wei2022,Oh2020,Herrmann2022,Gong2024}. Many of these have demonstrated significant learning capabilities. Recent advances in quantum hardware, such as trapped ions~\cite{Bruzewicz2019,Pino2021,Bermudez2017}, superconducting qubits~\cite{Arute2019,Kjaergaard2020,Krantz2019}, and photonic quantum computing~\cite{Slussarenko2019,Romero2024,Kok2007} have further fueled interest in practical implementations of QNNs.

However, the emergence of any quantum advantage by comparison with classical NNs is often not straightforward. With the limited capabilities of current quantum computers, it is not possible to compare a large classical NN with a similarly large QNN. Differences in activation functions, loss functions, and optimization algorithms are additional confounding factors. In this paper we examine, through classical simulation, the advantages of using quantum resources in neural networks by studying two natural quantizations of a standard multi-layer perceptron network~\cite{Kruse2022}. These approaches allow us to tune the network continuously between classical and quantum  modes of operation, allowing us to determine the benefit of quantum effects without making significant changes to the network's architecture.

\begin{figure}
    \centering
    \includegraphics[width=0.9\linewidth]{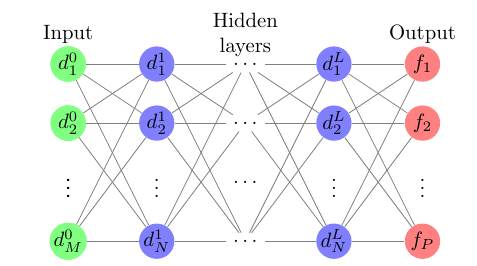}
    \caption{The classical binarized multi-layer perceptron network. It consists of an input layer of size $M$, $L$ hidden layers, each of size $N$, and an output layer.}
    \label{fig:nn_diag}
\end{figure}

The classical NN we will quantize is a binarized multi-layer perceptron~\cite{Yuan2023}, as shown in Fig.~\ref{fig:nn_diag}. The input layer has $M$ neurons. In a forward pass through the network these neurons will be initialized to the elements of a data vector $\mathbf d^0$. The network then contains $L$ hidden layers, each containing $N$ neurons, followed by an output layer of size $P$. The network is fully connected, each neuron is connected to all neurons in adjacent layers. The activation of the $i^\text{th}$ neuron in the $k^\text{th}$ hidden layer is
\begin{gather}
    d_i^k=\phi\left(\sum_jW_{ij}^{k-1}d_j^{k-1}\right),\\
    \phi(x)=\text{sgn}(x),
\end{gather}
where the $W^k$ are matrices of learnable weights and $\phi(x)$ is the activation function. The network is binarized in the sense that the activations of the hidden layers can only be $\pm 1$. This is in contrast to NNs which binarize only the weights~\cite{Courbariaux2015} or fully binarized NNs~\cite{courbariaux2016} which binarize both activations and weights. A similar binarized network was previously studied by two of the present authors in the context of spin model analogues~\cite{Barney2024}. The output vector is then computed as $\mathbf f=W^L\mathbf d^L$. This output vector is then the argument of some loss function that estimates the network's performance on some task.

For this binarized network, a slight change needs to be made the usual backpropagation approach of estimating the gradient of the loss function with respect to the weights~\cite{Linnainmaa1976,Rumelhart1987}. This is necessary because the first derivative of the activation function is either 0 or undefined everywhere. This can be overcome by using a clipped straight-through estimator for the gradient~\cite{bengio2013,courbariaux2016}. This amounts to treating the activation function as
\begin{equation}
    \phi_\text{back}(x)=\text{htanh}(x)=\begin{cases}
        x, & |x|\leq 1\\
        \text{sgn}(x), & |x|\geq 1
    \end{cases}
\end{equation}
when calculating the gradients.

\begin{figure}
    \centering
    \includegraphics[width=0.9\linewidth]{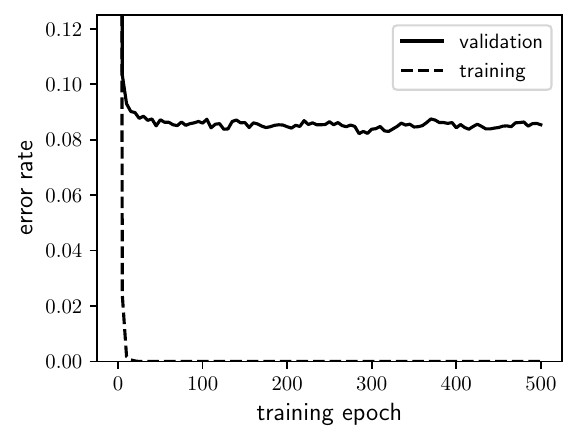}
    \caption{The validation and training error rates for the classical network as training progresses. The training error rate quickly vanishes while the validation error rate bottoms out at a nonzero value. This is an indication of overfitting.}
    \label{fig:classical_overfitting}
\end{figure}

We evaluate the performance of this classical network on the task of classifying images from the MNIST dataset~\cite{Lecun1998}, which contains images of handwritten digits 0-9. The full MNIST dataset contains 60,000 training images and 10,000 images for validation. In Fig.~\ref{fig:classical_overfitting} the training curves are shown for the classical network when trained on a subset of 5,000 MNIST training images. The relevant hyperparameters for this network are shown in Table~\ref{tab:hyperparams}. We see that, for this training set, overfitting~\cite{Ying2019} is very apparent in the network. The training error rate quickly vanishes, while the validation error rate bottoms out. This indicates that the model has become trapped in a minimum of the loss function that is unfavorable for generalization, a problem that is common when training on small datasets. In the remainder of this text we will examine how quantization may ameliorate this problem and regularize the network.

\begin{table}
    \centering
    \begin{tabular}{|c|c|}
        \hline
        \# hidden layers & 3\\
        \hline
        Hidden layer size & 512\\
        \hline
        Optimizer & Stochastic gradient descent\\
        \hline
        Loss function & Cross-entropy\\
        \hline
        Learning rate & 0.01\\
        \hline
        Momentum & 0.9\\
        \hline
        Batch size & 64\\
        \hline
        Training epochs & 500\\
        \hline
        Training dataset size & 5000\\
        \hline
        Validation dataset size & 10000\\
        \hline
    \end{tabular}
    \caption{The hyperparameters for the training and testing of the classical and quantum networks.}
    \label{tab:hyperparams}
\end{table}

\begin{figure}
        \centering
        \includegraphics[width=0.9\linewidth]{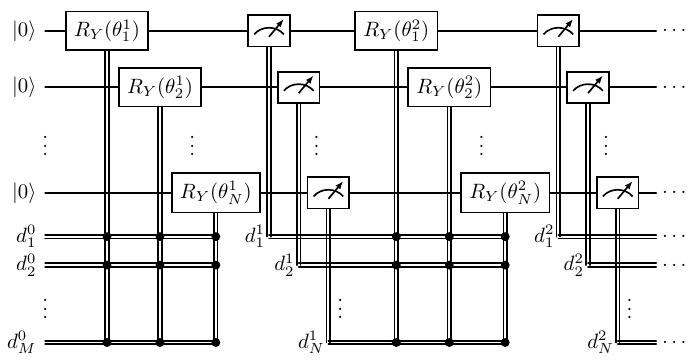}
        \caption{The quantum circuit implementing the hidden layers of the quantized network. Each step of the forward pass consists of rotating each qubit by an angle controlled by the classical channels, then measuring each qubit. These measurements are the activations which are then passed to the next layer of the network.}
        \label{fig:circuit_diag}
    \end{figure}

The quantized version of the classical NN shown in Fig.~\ref{fig:nn_diag} is implemented by replacing the hidden layers with a quantum circuit, as shown in Fig.~\ref{fig:circuit_diag}. The circuit is comprised of $N$ qubits and a set of classical channels. Initially, each of the qubits is set to the $|0\rangle$ state. The classical channels initially hold the input data vector $\mathbf d^0$. At each step in the forward pass, the $N$ qubits are individually rotated about the $y$-axis and then measured, with the measurement outcomes for the $k^\text{th}$ layer being the activations $\mathbf d^k$. The single-qubit rotation operator is
\begin{equation}
    R_Y(\theta)=e^{-i\frac\theta 2y},
\end{equation}
with $y$ being a Pauli operator. The rotation angle for the $i^\text{th}$ qubit in the $k^\text{th}$ layer is
\begin{gather}
    \theta_i^k=\frac\pi 2\begin{cases}
        1-\phi_a\left(\sum_jW_{ij}^{k-1}d_j^{k-1}\right), & k=1\\
        d_i^{k-1}-\phi_a\left(\sum_jW_{ij}^{k-1}d_j^{k-1}\right), & 2\leq k\leq L
    \end{cases},\\
    \phi_a(x)=\text{htanh}(x/a)=\begin{cases}
        x/a, & |x|\leq a\\
        \text{sgn}(x), & |x|\geq a
    \end{cases},
\end{gather}
where $\phi_a(x)$ is the network's activation function. The outputs $\mathbf{f}$ of the network are then calculated as in the classical network.

One can see that, when the limit $a\rightarrow 0$ is taken, the activation function reduces to $\phi_0(x)=\text{sgn}(x)$, as in the classical binarized network. In fact, in this limit, the rotation angles all become 0 or $\pm\pi$. One can verify by inspection that, if the output of the activation function is 1, the rotation operator will set the qubit to the $|0\rangle$ state. On the other hand, if the activation is $-1$, the qubit will be set to $|1\rangle$. Since the circuit only moves the qubits between computational basis states corresponding to the activations of the classical network, classical behavior is recovered in this limit.

Naturally, we are interested to see how model performance changes when tuning the parameter $a$ to nonzero values. Increasing $a$ causes the activation function to be ``stretched." This causes each rotation angle to no longer be constrained to be 0 or $\pm\pi$ when the magnitude of its preactivation is less than $a$, meaning that the rotations move the qubits into superpositions of the computational basis states. The outcomes of the projective measurements then become stochastic. In this way, the quantum resource of superposition is used to inject stochasticity into the network. Stochasticity is known to provide regularization in artificial NNs~\cite{Raiko2015,SietsmaA1991,Srivastava2014,bengio2013}.

\begin{figure}
        \centering
        \includegraphics[width=0.9\linewidth]{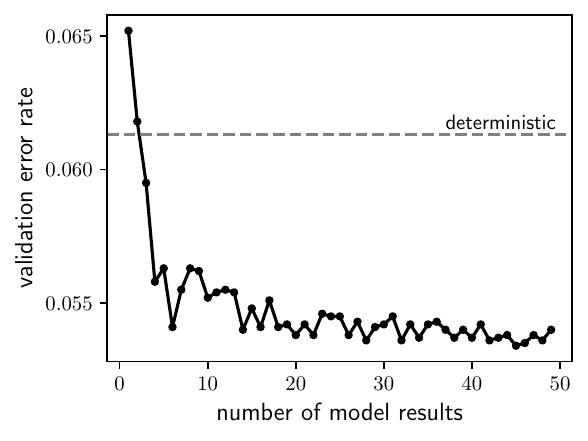}
        \caption{The validation error rate for a quantum network ($a=0.5$) after training as a function of the number of model results used to classify each image in the MNIST validation dataset.}
        \label{fig:mode_inference}
    \end{figure}

At this point, the question remains of how to best perform inference on the validation dataset using the weights obtained by training the quantum NN. There are two natural ways to do so. The first is to take the weights obtained and run inference deterministically. That is, set $a=0$ to recover classical behavior for the inference step. The second option we explore is to maintain the same value of $a$ for inference, but to pass each test datum through the network multiple times and select the most common prediction. Fig.~\ref{fig:mode_inference} shows how these two approaches compare for a particular trained network. We see that it only requires a few shots for quantum inference to outperform deterministic inference, though there are diminishing returns when taking more than approximately 10 shots. The validation error rates in the remainder of this text are computed by selecting the most common prediction from 15 shots for each image in the validation dataset.

\begin{figure}
        \centering
        \begin{tabular}{cl}
        \begin{minipage}[t]{0.05\linewidth} \centering (a) \end{minipage} & 
        \begin{minipage}{0.9\linewidth} 
            \includegraphics[width=\linewidth]{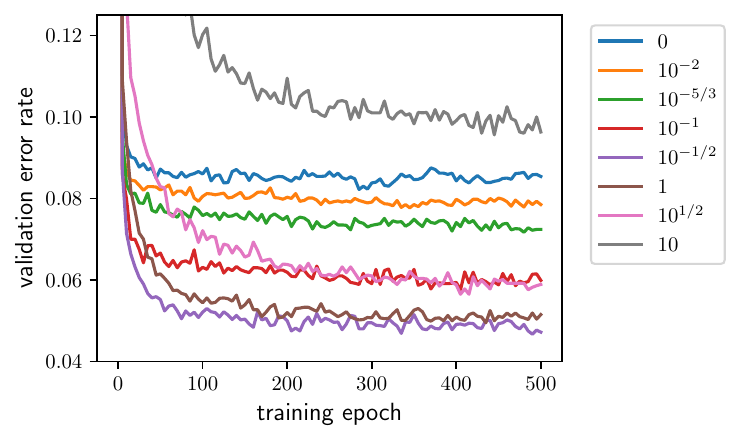}
        \end{minipage} \\
        \begin{minipage}[t]{0.05\linewidth} \centering (b) \end{minipage} & 
        \begin{minipage}{0.75\linewidth} 
            \includegraphics[width=\linewidth]{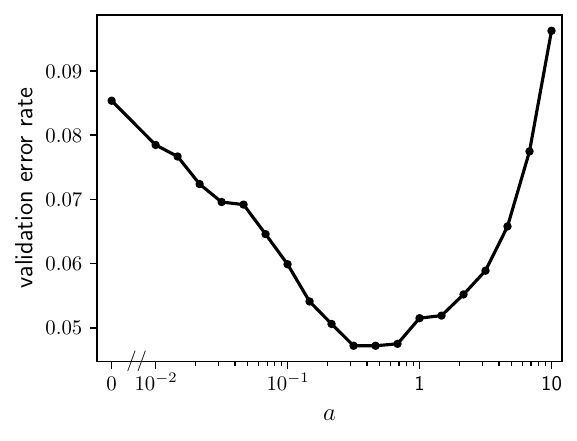}
        \end{minipage}
    \end{tabular}
        
        \caption{(a) Validation error rate curves as training progresses in the quantized network for selected values of $a$. (b) The validation error rate as a function of $a$. The best result is achieved for nonzero $a$, indicating that quantum effects improve performance.}
        \label{fig:averr}
    \end{figure}

In Fig.~\ref{fig:averr}(a) we see the training curves showing the validation error rate as training progresses for some selected values of $a$. We observe that the quantized network reaches significantly lower error rates than the classical network for some nonzero values of $a$. In Fig.~\ref{fig:averr}(b) we see the final validation error rates as a function of $a$. Our best result is obtained for $a=10^{-1/2}\approx0.316$, which achieves an error rate of 0.0472. Increasing $a$ further actually reduces performance until the quantum network performs worse than the classical one. This is expected, since introducing too much randomness to the network will inevitably impede its ability to learn.

We can understand the increase in performance for nonzero $a$ in terms of quantum tunneling. The objective of the learning algorithm is to descend down the landscape of the loss function to find a minimum. Doing this purely deterministically can cause the model to become trapped in a local minimum that leads to good performance on the training set, but has poor generalization performance on the validation dataset. This issue becomes more common when training on small datasets. By adding quantum randomness to the model, we make it easier for the model to tunnel out of an unfavorable local minimum and find a more favorable minimum that leads to better generalization.

\begin{figure}
    \centering
    \includegraphics[width=0.9\linewidth]{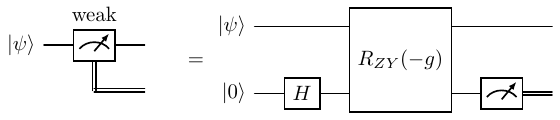}
    \caption{The quantum circuit that implements a weak measurement of a neuron qubit by entangling with and then measuring an ancilla qubit.}
    \label{fig:wm_circuit}
\end{figure}

In addition to the quantization describe above, we examined another method of quantization by way of weak measurements. This may be done by setting $a=0$ in the previous treatment and replacing the projective mid-circuit measurements with weak measurements. Each of these weak measurements is implemented by entangling each neuron qubit with an ancilla qubit that is prepared in an equal superposition of the computational basis states, and then projectively measuring the ancilla qubit. The entanglement is effected by the two-qubit rotation gate
\begin{equation}
    R_{ZY}(-g)=e^{i\frac g 2z\otimes y_\text{anc}},
\end{equation}
where $g$ is a parameter that controls the strength of the entanglement. Fig.~\ref{fig:wm_circuit} shows the circuit diagram for this weak measurement protocol.

If the neuron qubit is initially in the state $|\psi_0\rangle=\alpha|0\rangle+\beta|1\rangle$, then, after the entangling gate, the two-qubit system will be in the state
\begin{multline}
    |\Psi\rangle=\sqrt{(1+\sin g)/2}(\alpha|00\rangle+\beta|11\rangle)\\
    +\sqrt{(1-\sin g)/2}(\alpha|01\rangle+\beta|10\rangle).
\end{multline}
    
We need only consider $g$ within the interval $[0,\frac\pi 2]$. When $g=0$, the two-qubit system can be factored into a product state; no entanglement is created. On the other hand, when $g=\frac\pi 2$, the system is in a maximally entangled Bell state~\cite{Bell1964}; the ancilla measurement effectively projectively measures the neuron qubit, recovering classical behavior. Measuring the ancilla qubit will, in general, disturb the neuron qubit. If the measurement outcome is $d$, the state of the neuron qubit will become
\begin{equation}
    |\psi_f\rangle=\alpha\sqrt\frac{1+d\sin g}{1+d\langle z\rangle\sin g}|0\rangle+\beta\sqrt\frac{1-d\sin g}{1+d\langle z\rangle\sin g}|1\rangle,
\end{equation}
where $\langle z\rangle=|\alpha|^2-|\beta|^2$ is the expectation value of a $z$ measurement of the neuron qubit before the ancilla measurement.

Since the activations of the network are now determined by measuring the ancilla qubits, each activation has the probability $\frac 1 2(1-\sin g)$ to be ``wrong" in comparison to what the classical activation would be. This approach provides an alternative way to introduce stochasticity to the network. The nature of the stochasticity introduced by weak measurements is different than the previous approach because each activation has the same probability of being ``wrong." In contrast, the previous approach only has a nonzero probability for an activation to differ from the classical activation when the preactivation has a magnitude less than $a$. In this way, neurons with large preactivations, indicating more certainty, are unaffected by quantum randomness. The advantage of the weak measurement approach is that quantum coherence is not fully destroyed at each step in the forward pass.

\begin{figure}
    \centering
    \begin{tabular}{cl}
        \begin{minipage}[t]{0.05\linewidth} \centering (a) \end{minipage} & 
        \begin{minipage}{0.9\linewidth} 
            \includegraphics[width=\linewidth]{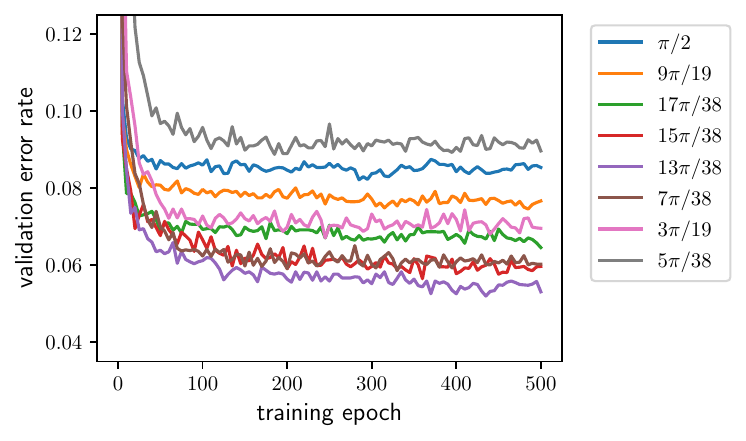}
        \end{minipage} \\
        \begin{minipage}[t]{0.05\linewidth} \centering (b) \end{minipage} & 
        \begin{minipage}{0.75\linewidth} 
            \includegraphics[width=\linewidth]{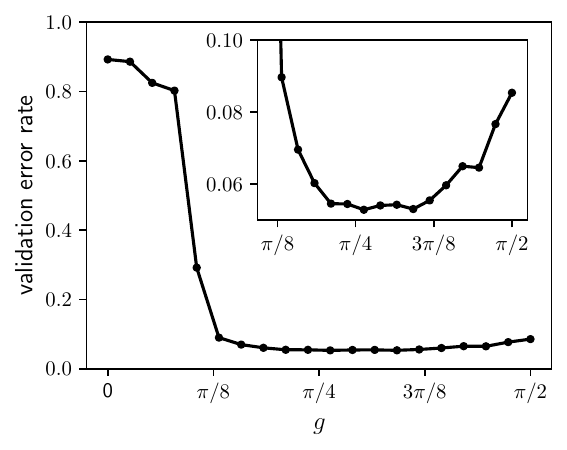}
        \end{minipage}
    \end{tabular}
    \caption{(a) Validation error rate curves as training progresses in the quantized network with weak measurements for selected values of $g$. (b) The validation error rate as a function of $g$. The inset is the same but zoomed in near the minimum of the curve. The best result is achieved for $g<\frac\pi 2$, indicating that quantum effects improve performance.}
    \label{fig:gverr}
\end{figure}

In Fig.~\ref{fig:gverr}(a) we see the training curves showing the validation error rate as training progresses for selected values of $g$. Similar to our previous approach, we see that convergence time is mostly unaffected until $g$ becomes sufficiently small. Before that point, there are values of $g$ for which the model performs significantly better than the classical network. In Fig.~\ref{fig:gverr}(b) we see the final validation error rates as a function of $g$. We see that, for $g$ less than approximately $\frac\pi 8$, the network quickly loses its ability to learn at all. Our best result was obtained when $g=\frac{5\pi}{19}$, which achieved a validation error rate of 0.0529. This is slightly worse than the best performing model using our previous approach. The lack of discrimination between neurons with large or small preactivations in the weak measurement approach is one possible explanation for this, though the difference is small enough that its significance is uncertain.

\begin{figure}
    \centering
    \includegraphics[width=0.9\linewidth]{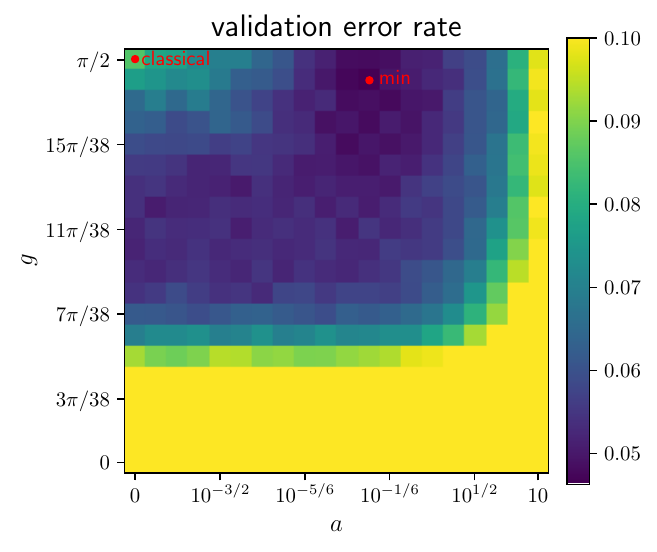}
    \caption{Validation error rates results for networks making use of both quantization approaches, varying both $a$ and $g$.}
    \label{fig:agverr}
\end{figure}

It is also natural to examine model performance when our two approaches are combined. That is, to quantize the model by varying both $a$ and $g$. The results of this combined approach are shown in Fig.~\ref{fig:agverr}. The classical result is shown in the top left corner, where $a=0$ and $g=\frac\pi 2$. We see that performance is improved by increasing $a$, decreasing $g$, or both simultaneously, up to a point. The best performance we observed was when $a=10^{-1/3}\approx 0.464$ and $g=\frac{9\pi}{19}$, which achieved a validation error rate of 0.0463, though performance does not vary much along the curved valley floor shown in Fig.~\ref{fig:agverr}.

In this paper we explored two approaches to quantizing a binarized multi-layer perceptron model. The first approach uses single qubit rotations to create quantum superpositions for neurons with sufficiently small pre-activations. Each qubit is then projectively measured, leading to stochastic activations for these neurons. The second approach entangles the neuron qubits with ancilla qubits. The activations are then found by projectively measuring the ancilla qubits. This approach is independent of the pre-activation magnitudes, but has the advantage of not completely destroying coherence at each step in the circuit. Both approaches use quantum resources to introduce stochasticity to the network, and both allow for continuous tuning between classical and more quantum modes of operation. We find that, when training on a reduced MNIST dataset, introducing quantum stochasticity leads to a significant reduction in the validation error rate. This improvement can be understood as an enhancement in the model's ability to tunnel away from an unfavorable local minimum of the loss function to a more favorable minimum which leads to better generalization.

In this work we were not primarily concerned with creating state-of-the-art QNNs, instead seeking to understand how performance changes as we move continuously between classical and quantum modes of operation without changing the model architecture. We consider the quantized networks we examined to be a starting point in exploring how quantum resources can be utilized to improve model performance. There are many possible quantum advantages that remain to be explored within our framework, such as representing data more efficiently with multi-qubit superposition states, quantum algorithms for optimizing the loss function, and exploring connections to measurement induced phase transitions~\cite{Skinner2019}. Thus far, we have only explored the behavior of these networks when trained on classical data. It would be interesting to see what advantages they might have when being trained on data collected from experiments on physical quantum systems. Currently, we are exploring generalizations of the architectures in this paper which entangle the neuron qubits through multi-qubit gates. Additionally, we are working towards implementing our current architectures on actual quantum hardware, extending the results of this paper beyond classical simulation.

We thank Norbert M. Linke and Alaina M. Green for helpful conversations on the topic of practical quantum computation. This research was sponsored by the U.S. Army Research Office under Grant Number W911NF-23-1-0241 and the National Science Foundation under Grant No. DMR-203715.

\bibliography{paper_bib}
\end{document}